\documentstyle[twoside,fleqn,espcrc2]{article}

\newcommand{\AmS}{{\protect\the\textfont2
  A\kern-.1667em\lower.5ex\hbox{M}\kern-.125emS}}

\title{Atmospheric neutrino interactions in Soudan-2
 \thanks{Presented at TAUP97, the 5th International 
Workshop on Topics in Astroparticle and Underground Physics, Sept. 7-11,
1997, Laboratori Nazionali del Gran Sasso, Assergi, Italy.}}

\author{T. Kafka \thanks{High Energy Physics, Tufts University,
4 Colby St., Medford, MA 02155, USA}
\vskip 0.2cm
\noindent {\it for the Soudan-2 Collaboration}
\vskip 0.1cm
\noindent (Argonne National Laboratory, University of Minnesota, Tufts
University, USA; Oxford University, Rutherford Appleton Laboratory, UK)}

\begin{document}

\begin{abstract}
A new measurement of the atmospheric $\nu_\mu/\nu_e$ ratio-of-ratios,
$0.61\pm 0.15\pm 0.05$, has been obtained using a 3.2-kty exposure
of the Soudan-2 underground detector.
This measurement, based upon neutrino reactions in an iron tracking calorimeter
of honeycomb-lattice geometry, is in
agreement with the anomalously low value reported by the
underground water detectors.

\end{abstract}

\maketitle

\section{INTRODUCTION}

%\subsection{Spacing}

Soudan-2 is a fine-grained iron tracking calorimeter located 2100 mwe
underground in Soudan, Minnesota, USA. The experiment 
has been taking data since 1989 when the detector was one
 quarter of its full size; the construction was completed at the end of 1993.
The experiment continues to take data with 90\% live time, having accumulated
3.6 kty up to the date of this Conference.
The first result on the ratio of ratios of the atmospheric neutrinos,
\begin{equation}
 R = {{[(\nu_\mu+\bar\nu_\mu)/(\nu_e+\bar\nu_e)]_{Data}}
 \over{[(\nu_\mu+\bar\nu_\mu)/(\nu_e+\bar\nu_e)]_{MC}}} \label{eq:R}
\end{equation}
was published earlier this year.\cite{First}

\section{THE DETECTOR}

The two main components of the Soudan-2 apparatus are the Central Detector
and the Active Shield.
 The {\it Central Detector} consists of 224 $1\times1\times2.5$ m calorimeter
modules weighing 4.3 tons each. The bulk of the mass consists of 1.6-mm-thick
corrugated steel sheets which are interleaved with drift tubes.
Each half of a 1.5-cm-diam., 1-m-long, resistive drift tube is read out
by a crossed pair of a vertical anode wire and a horizontal cathode pad
that determine two coordinates of a tube crossing;
the third coordinate is deduced
from the drift time. Readout from the 3.4 million half-tubes is accomplished
using 136,000 anode and cathode channels that are multiplexed
before the signals are digitized.

The {\it Active Shield} covers the walls of the Soudan-2 cavern enclosing the
Central Detector as hermetically as possible. Only the calorimeter's
support columns must pass through the shield,
 adding up to about 5 m$^2$, out of the total shield area
of 1700 m$^2$. The shield consists of 1529 modules with a typical size of
19 cm $\times$ 7 m, each containing a double layer of proportional tubes
with a digital readout. A coincidence of two adjacent layers is required for 
a hit in the off-line analysis.

Detailed descriptions have been published
of the design and performance of Soudan-2 calorimeter modules
\cite{NIM}, and of the Active Shield system \cite{NIM_VS}.

\section{CONTAINED EVENTS}

The contained events (CEV) in an underground detector are expected to be
atmospheric neutrino interactions, where the neutrinos originate in decays of
mesons and muons in cosmic ray showers in the atmosphere.
It is these interactions that we model in our Monte Carlo simulation
 and compare with data.
 We use the Bartol atmospheric neutrino flux \cite{Bartol}, and
 every MC event is overlayed onto a random pulser event
to acquire a realistic noise pattern.
 In the following, we used simulated events corresponding
to an 18.3 kty exposure of Soudan-2; the dependence of the neutrino
flux on the solar cycle was taken into account.

During the period April 1989 through December 1996,
we analyzed 75 million triggers, most of which are initiated by
throughgoing muons or ``noise" (both electronic noise and triggers due to
local radioactivity in the cavern). 75\% of the Monte Carlo events pass the
hardware trigger requirement.

All triggers are processed by a two-stage software filter.
The software performs a number of data quality checks, and rejects events
that conform to one of the identified noise patterns. To check for event
containment, the software requires that
{\it (i)} no part of the event is within 20 cm of the detector surface,
and that {\it (ii)}
no track is located and oriented in such a way that it could
enter the calorimeter undetected (e.g. through a crack between modules).
The number of events satisfying the software requirements
is 15,000 per kty. About 50\% of Monte Carlo events that satisfy the trigger
conditions pass the software filter.

\par
The last step in the data reduction chain is a two-stage physicist scan.
The resulting CEV sample contains typically 500 events per kty, representing
an overall data reduction of $2 \times 10^{-5}$. 
The Monte Carlo events are injected into the data stream before the scanning
stage so that physicists scan data and MC simultaneously without knowing
which is which. Of the remaining Monte Carlo events, 70\% pass the scan
requirements, indicating an overall neutrino CEV detection efficiency of
 about 26\%.

\begin{table}[t]
% space before first and after last column: 1.5pc
% space between columns: 3.0pc (twice the above)
%\setlength{\tabcolsep}{1.5pc}
\setlength{\tabcolsep}{0.6pc}
% -----------------------------------------------------
% adapted from TeX book, p. 241
\newlength{\digitwidth} \settowidth{\digitwidth}{\rm 0}
\catcode`?=\active \def?{\kern\digitwidth}
% -----------------------------------------------------
\caption{The (mis)identification matrix of the atmospheric neutrino
Monte Carlo events. The numbers are percentages;
every column adds to 100\%.}
\label{tab:confusion}
%\begin{tabular}{\textwidth}{@{}l@{\extracolsep{\fill}}cccc}
%\footnotesize{
%\begin{tabular}{lcccc}
\begin{tabular}{@{}l@{\extracolsep{\fill}}rrrrr}
\hline
Reaction & & Track & Shower & Multi & Proton \\
\hline
$\nu_\mu$ CC & & 87.2   &  2.7 & 37.3 & 28.0 \\
$\nu_e$ CC   & &  4.3   & 92.8 & 41.6 &  8.8 \\
NC           & &  8.5   &  4.5 & 21.1 & 63.2 \\
\hline
\end{tabular}
\end{table}

In the last scanning stage, all events are classified into one of three
topologies: Tracks, showers, and multiprongs. Among the neutrino events, 
most of the single prong topologies are quasielastic interactions,
$$ \nu_\l + n \rightarrow \l^- + p ,$$
and
$$ \bar\nu_\l + p \rightarrow \l^+ + n ,$$
where $\l = \mu, e$. Presence of a proton at the primary vertex is not
included in the classification. Single track events are subject to an
additional scrutiny: All tracks that pass criteria of straightness
and heavy energy deposition are removed from the track sample and
reclassified as lone protons. For single shower events, we require that
they have nine or more hits
to remove a range where, at present, we are less confident in
separating signal from certain kinds of noise in the detector.
 Examination of Monte Carlo neutrino events shows the degree of success
of our classification, as summarized in
 Table~\ref{tab:confusion}. We see that most tracks and showers indeed have
$\nu_\mu$ and $\nu_e$ flavor, respectively, as expected. Multiprongs
include interactions of all neutrino flavors as well as neutral current
(NC) events, and special selection is needed to determine their neutrino flavor.
They are not included in the results presented here.

\textfloatsep 24pt
\begin{figure}[t]
\vspace {150.0pt}
\includegraphics{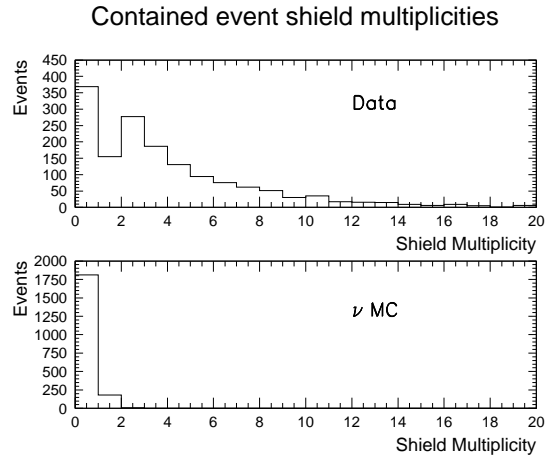}
\caption{Active Shield hit multiplicities for (a) data, (b) Monte Carlo.}
\label{fig:vsmult}
\end{figure}

In the CEV analysis, the Active Shield plays a crucial role.
Figure~\ref{fig:vsmult} displays the shield hit multiplicity distribution
for the data and for the Monte Carlo events.
Among Soudan-2 contained events,
80\% have one or more associated in-time shield hits.
In the terminology used below, events without any shield hits are called
{\it Gold events}, while events with two or more shield hits are called
{\it Rock events}. 
The former are mostly neutrino interactions, while the latter are
due to neutral hadrons or photons created in an inelastic interaction
of a cosmic-ray muon in the rock in the vicinity of the Soudan-2 cavern.
Events with one shield hit are a mixture of the two - they contain
neutrino events associated with a random shield hit (see the Monte Carlo
distribution in Fig.~\ref{fig:vsmult}) as well as events originating in the
rock. We do not consider these events in our further analysis.
Our task is now to ascertain if there are any non-neutrino events in the
Gold Sample.

\section{NON-NEUTRINO BACKGROUND}

The Gold Sample is expected to contain two classes of events;
\begin{itemize}
\item [-] neutrino interactions; and
\item [-] possible background due to interactions of particles originating
in the rock, for one of two reasons: {\it (i)} there are charged particles
originating in the rock passing through the Active Shield, but there are
no hits due to shield inefficiency (a 5\% effect); or {\it (ii)}
there are no charged particles passing from the rock through the shield,
called {\it No-shield Rock events}.
\end{itemize}

Rock events include gamma conversions creating showers and
neutron interactions creating tracks, showers, and multiprongs.
We note that the density  of the Soudan-2 Central Detector medium is
1.6 g/cm$^3$, the radiation length
is 9.7 cm, and the nuclear interaction length is 81 cm.
We therefore expect differences between vertex distributions of neutrino and
Rock events, the Rock event distribution exhibiting a noticeable attenuation
as a function of depth in the detector. We define
{\it Penetration Depth} to be
the shortest distance from the event vertex to the detector exterior
(excluding the floor). Such definition is necessary as we do not know
the direction of the incoming particle in most of the events.

\begin{figure}[hbt]
\vspace {220.0pt}
\includegraphics{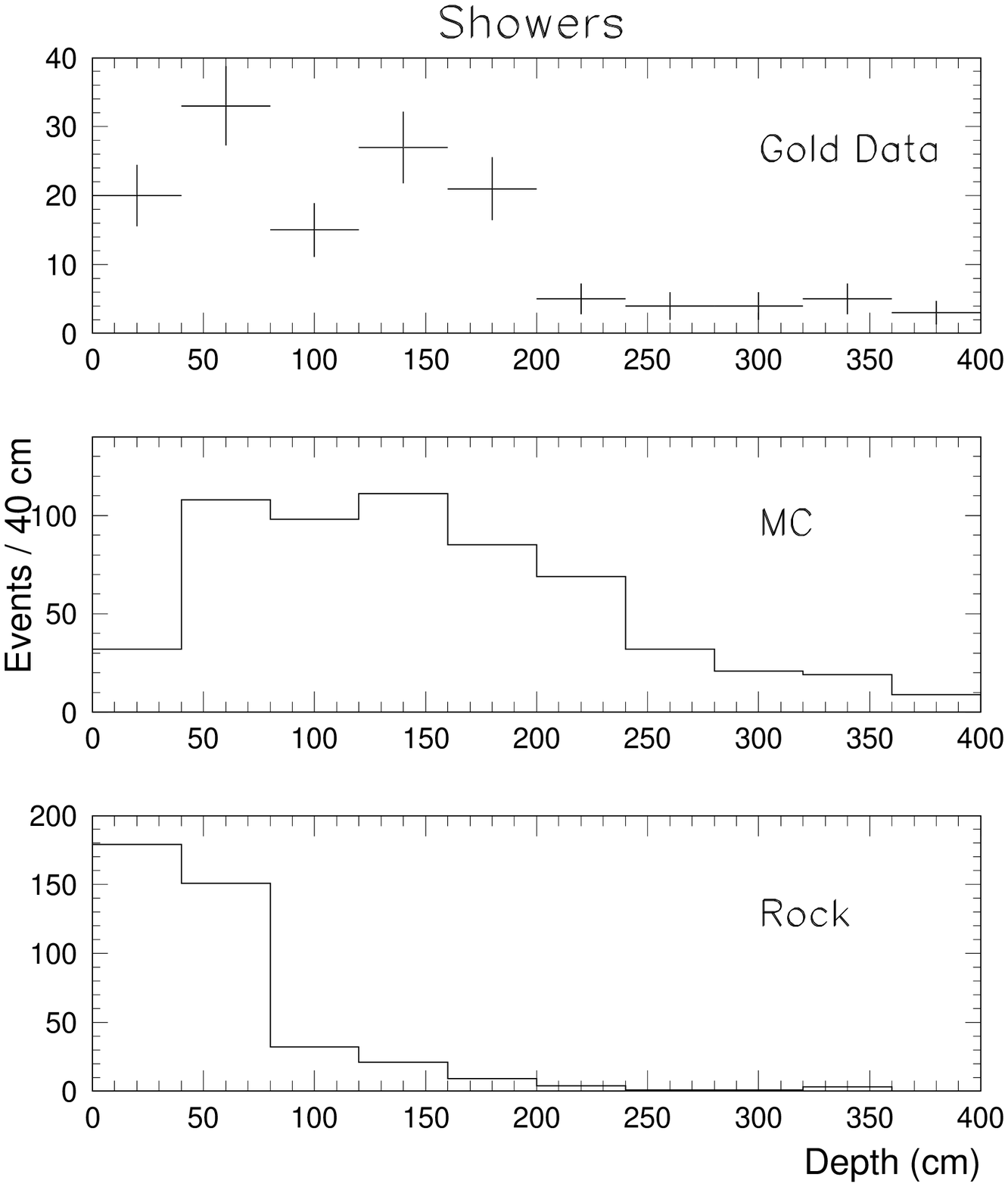}
\caption{Penetration Depth distributions for single shower evens:
 (a) Gold data sample, (b) Monte Carlo events, and (c) Rock Events.}
\label{fig:depth_s}
\end{figure}

\begin{figure}[hbt]
\vspace {220.0pt}
\includegraphics{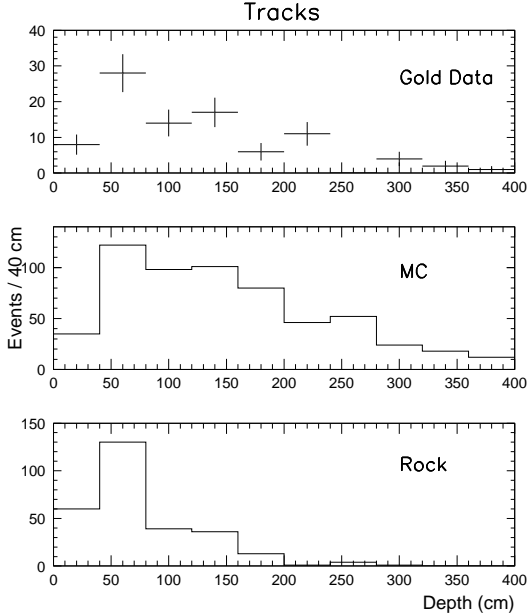}
\caption{Penetration Depth distributions for single track events:
 (a) Gold data sample, (b) Monte Carlo events, and (c) Rock events.}
\label{fig:depth_t}
\end{figure}

The penetration depth distributions are shown in Fig.~\ref{fig:depth_s} for
single shower events and in Fig.~\ref{fig:depth_t} for single track events.
 In both figures we
observe that the Monte Carlo distributions extend over the whole range
of the penetration depth, while the Rock distributions are confined
to smaller depths. The Gold Data distributions do extend over the whole
range, indicating presence of neutrino events, but it is not easy to
separate the two components by eye. A hint is perhaps the step down from the
second to the third bin which is present in the Gold Data and Rock
 distributions, but not in the Monte Carlo histogram.

We therefore adopt the following procedure to determine the amount of
non-neutrino background in the Gold Data: We fit the Gold Data depth
distributions to the sum of the expected neutrino (MC) and 
background (Rock) depth distributions. We choose the Extended
Maximum Likelihood Method, a binned maximum likelihood which
is appropriate for low statistics distributions, which preserves the
total number of events in the distributions being fitted, and which takes
into account finite statistics of the fit components, the Monte Carlo
and Rock distributions.

\begin{figure}[hbt]
\vspace {220.0pt}
\includegraphics{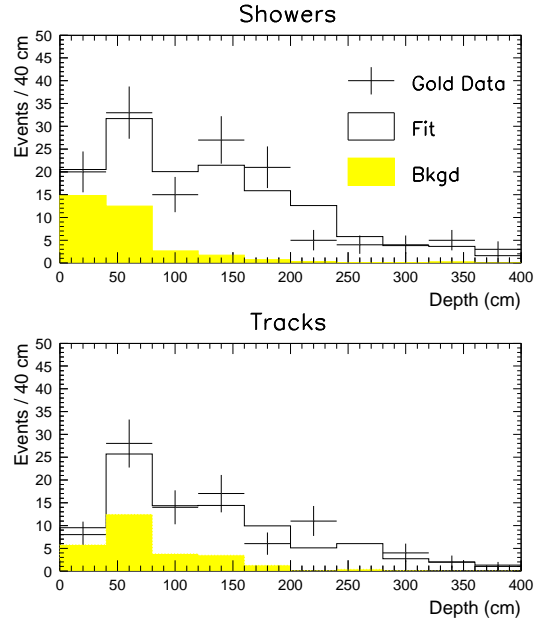}
\caption{Fit to the Penetration Depth distributions for (a) showers, and
(b) tracks.}
\label{fig:fit}
\end{figure}

 The result is displayed in Fig.~\ref{fig:fit}. Here the Gold Data are depicted
by the crosses, the solid-line histogram is the fit, and the shaded areas
represent the amount of the Rock related background.

\section{ATMOSPHERIC NEUTRINO FLAVOR RESULTS}

%Our final samples are summarized in Table~\ref{tab:results}. We see that
Our final samples are summarized in Table 2. We see that
both track and shower samples have 25 - 30\% non-neutrino background.
The corrected number of neutrino showers agree with the number of Monte
Carlo showers while the corrected number of neutrino tracks exhibits
a deficit when compared to Monte Carlo.

\begin{table}[hbt]
\label{tab:results}
\caption{Soudan-2 results for 3.2 kty.}
\setlength{\tabcolsep}{0.7pc}
%\begin{tabular}{lcc}
\begin{tabular}{@{}l@{\extracolsep{\fill}}ccc}
\hline
& & Tracks & Showers\\
\hline
 Gold, total & & 91 & 137 \\
 Rock        & & 284 & 401 \\
Bkgd in Gold & & 27.1$\pm$11.4 & 33.2$\pm$9.9 \\
(fraction of Gold & & 29.8\% & 24.2\%) \\
 Monte Carlo & & 588 & 584 \\
 (norm. to 3.2 kty & & 101.4$\pm$4.2 & 100.7$\pm$4.2) \\
$\nu$ in Gold & & 63.9$\pm$13.1 & 103.8$\pm$13.6 \\
\hline
\end{tabular}
\end{table}

In order to measure the $\nu$ flavor composition in as unbiased a way as
possible, it is customary to use a ``ratio of ratios". We express
$R$ defined in Eq.~\ref{eq:R} using
our number of tracks, $T$, as a measure of $\nu_\mu$ events, and
number of showers, $S$, as a measure of $\nu_e$ events:
\begin{displaymath}
 R = {{T_\nu/S_\nu}\over{T_{MC}/S_{MC}}} .
\end{displaymath}
Before the background correction, we obtain $R = 0.66 \pm 0.10$ using the raw
Gold sample counts. Using the corrected numbers of neutrino events
in our Gold sample from Table 2, we obtain
\begin{displaymath}
 R = 0.61 \pm 0.15 \pm 0.05 ,
\end{displaymath}
where the first error is statistical, and the second is
systematic.

Our value of $R$ is consistent with results of underground water detectors
reviewed in Ref.~\cite{Kearns}.
The indication is that there are too few single tracks, i.e.
too few $\nu_\mu$ and $\bar\nu_\mu$ events in the
underground neutrino flux, usually explained as due to neutrino
oscillations.

\begin{figure}[hbt]
\vspace {220.0pt}
\includegraphics{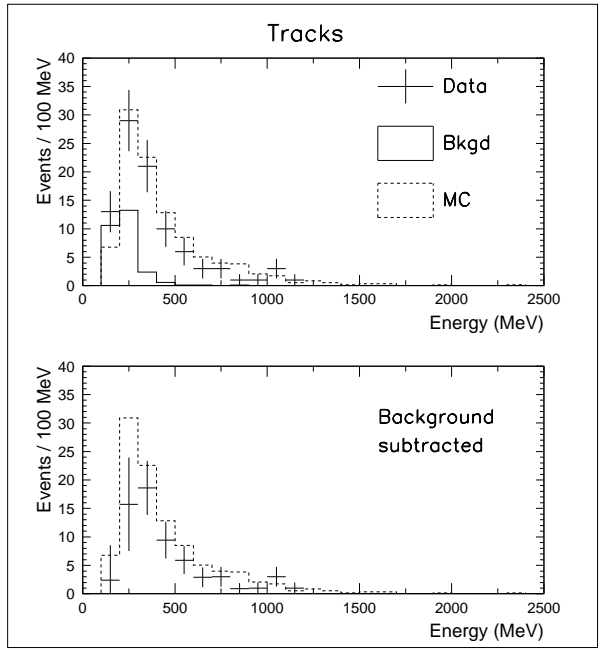}
\caption{Lepton energy distribution in the Gold Track sample.}
\label{fig:energy_t}
\end{figure}

\begin{figure}[hbt]
\vspace {220.0pt}
\includegraphics{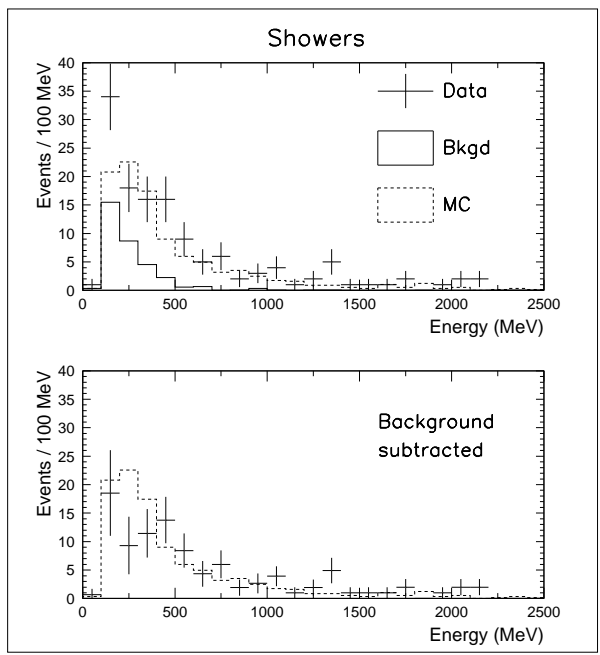}
\caption{Lepton energy distribution in the Gold Shower sample.}
\label{fig:energy_s}
\end{figure}

Finally, we present lepton energy distributions for our track sample in
Fig.~\ref{fig:energy_t} and for our shower sample in 
Fig.~\ref{fig:energy_s}. In the upper portion of the figures,
 the crosses represent our Gold Data before any correction, and the solid
histogram shows the Rock background. The dashed histograms are
the Monte Carlo distributions. In the lower portion of the figures
we show our corrected data distributions. We note that the background
is confined not only to smaller penetration depths, but also to smaller
energies than the neutrino event distributions.

Our plans for the near future include the extraction of the
neutrino flavor ratio from the multiprong events, and neutrino
oscillation analysis of our data.

\end{document}